\documentclass{emulateapj}

\usepackage{amsmath}
\usepackage{verbatim}
\usepackage{wasysym}
\usepackage{multirow}
\usepackage{soul} 
\usepackage[dvips]{color}

\newcommand\addtag{\refstepcounter{equation}\tag{\theequation}}

\usepackage{times}
\usepackage{textcomp}

\usepackage{aas_macros}

\usepackage{booktabs}

\begin{document}

\title{Detection Efficiency of Asteroid Surveys}

\author{Pasquale Tricarico}
\affil{Planetary Science Institute \\ 1700 East Fort Lowell, Suite 106, Tucson, AZ 85719, USA}
\email{\href{mailto:tricaric@psi.edu}{e-mail: tricaric@psi.edu}}

\begin{abstract}
A comprehensive characterization of the detection efficiency of nine of the major asteroid surveys 
that have been active over the past two decades is presented.
The detection efficiency is estimated on a nightly basis
by comparing the detected asteroids with the complete catalog of known asteroids propagated
to the same observing epoch.
Results include a nightly estimate of the detection efficiency curves 
as a function of apparent magnitude and apparent velocity of the asteroids,
as well as a cumulative analysis to estimate the overall performance of each survey.
The limiting magnitude distribution is estimated for each survey,
and it is then modeled as a function of telescope aperture
to obtain an estimate over a wide range of apertures.
\end{abstract}


\maketitle

\section{Introduction}

Asteroids are small bodies orbiting the Sun,
scattered across the Solar System with a large majority 
concentrated in the main belt region between the orbits of Mars and Jupiter.
Their orbital distribution and physical properties
are a primary source of information for reconstructing the processes which led to
the formation and evolution of the Solar System
\citep{1969JGR....74.2531D,2005Icar..175..111B}.
While the first asteroids were discovered over two centuries ago,
most of the known asteroids have been discovered by dedicated surveys over the past two decades.
These modern asteroid surveys have been tracking asteroids
using robotic telescopes, charge-coupled device (CCD) detectors, and dedicated software
that typically automates the bulk of the data acquisition and reduction while 
leaving the final decision on each detection to expert human observers.
The ability to produce unbiased estimates of the population of asteroids
relies in large part on accurately assessing the detection efficiency
of the asteroid surveys
\citep{1998Icar..131..245J}.
This is the main motivation to perform a comprehensive characterization
of the modern asteroid surveys active over the past two decades.
Additionally, surveys can take advantage of this comparative analysis to assess the
tradeoff of different observing strategies and setups.

The detection efficiency is an important diagnostic,
and surveys have adopted different approaches over time in order to determine it.
In \cite{1999AJ....117.1616P} 
the detection efficiency of selected clear nights is determined for the
Near-Earth Asteroid Tracking (NEAT) survey,
by comparing the detected objects with the list of known objects,
both as a function of apparent magnitude and apparent velocity (angular rate) of the asteroids.
In \cite{LINEAR2003} the detection efficiency of individual fields near the ecliptic plane is determined for the 
Lincoln Near-Earth Asteroid Research (LINEAR) survey,
also by comparing the detected objects with the list of known objects,
but only as a function of apparent visual magnitude.
The binned data are then fitted to a sigmoid curve.
In \cite{Jedicke_CSS} a similar approach is followed
to determine the detection efficiency of the Catalina Sky Survey.
When estimating the detection efficiency of telescopes
with larger apertures and smaller fields of view,
both \cite{2007P&SS...55.1113Y} and \cite{2009Icar..202..104G}
injected synthetic asteroids in their acquired fields
in place of the insufficient number of known objects in the fields.
In \cite{2011ApJ...743..156M} the detection efficiency for the NEOWISE survey
was determined by comparing detections with known objects in the field of view,
as a function of the infrared magnitude at 12 and 22 $\mu$m wavelengths.

In this study we have included all available data from nine of the major asteroid surveys
that have been active over the past two decades,
all of which are ground-based except NEOWISE,
see Table~\ref{tab:survey_properties}
and \cite{2015arXiv150304272J} for a recent review.
Available data from the beginning of operations of each survey through August 2014 are included.
The survey selection is dictated principally by the availability of a sufficient volume of data
to allow a statistical treatment,
such as a sufficient number of observations reported on each single night
to reliably determine the detection efficiency parameters for that night,
and a large enough number of nights to allow 
a characterization of the overall survey performance.

\begin{deluxetable}{ccccccrr}
\tablewidth{\columnwidth}
\tablecaption{Surveys Properties}
\tablehead{ \colhead{Name} & \colhead{MPC} & \colhead{$D$} & \colhead{$f/D$} & \colhead{$s$} & \colhead{FOV} & \multicolumn{2}{c}{\emph{nights}}}
\startdata
NEAT                 & 644 & 1.2 & 3.0   & 1.0 & 5.0 & 1031 &  431 \\
Spacewatch           & 691 & 0.9 & 3.0   & 1.0 & 2.9 & 1607 & 1585 \\
LONEOS               & 699 & 0.6 & 1.8   & 2.6 & 8.0 & 1328 &  978 \\
Catalina Sky Survey  & 703 & 0.7 & 1.8   & 2.5 & 8.2 & 1954 & 1729 \\
LINEAR               & 704 & 1.0 & 2.2   & 2.3 & 2.0 & 2852 & 2187 \\
NEOWISE              & C51 & 0.4 & 3.4   & 2.8 & 0.6 &  191 &  188 \\
Siding Spring Survey & E12 & 0.5 & 3.5   & 1.8 & 4.2 & 1931 &  806 \\
Pan-STARRS           & F51 & 1.8 & 4.0   & 0.3 & 7.0 &  896 &  660 \\
Mt.~Lemmon Survey    & G96 & 1.5 & 2.0   & 1.0 & 1.2 & 1629 & 1469 \\
\enddata
\tablecomments{Columns:
survey name and MPC code,
aperture $D$ in meters,
focal ratio $f/D$,
pixel scale $s$ in arcsec/pixel,
and field of view (FOV) in deg$^2$.
The two nights entries are the total number of nights available (left),
and the number for which a good fit was obtained (right).
Entries are sorted by MPC code.
}
\label{tab:survey_properties}
\end{deluxetable}

It is important to clarify that here by ``detection" we mean that 
observations have been fully processed by a survey, 
then reported to the Minor Planet Center (MPC),
and finally linked to a known asteroid.
There may be instances in which observations of a potential asteroid
are not linked to a know asteroid, neither they lead to constrain sufficiently well
the orbit of a new asteroid.
These so-called \emph{one night stands} are not considered detections 
until they are linked to known asteroids.

\section{Methods}

This analysis includes several surveys and thus requires us 
to use a method to determine the detection efficiency that is generic enough to only require publicly available data,
in order to be as inclusive as possible and to have the highest possible statistical significance.
Data from each survey are analyzed on a nightly basis,
which implies that the varying observing conditions and related possible detection efficiency changes during the night
are averaged into a single estimated detection efficiency for that night.
This choice is dictated by the goal of using a sufficient amount of data 
that can be achieved when using a full night of data, but would not be possible if using shorter time intervals.

All the input data are from the MPC:
the asteroid observations are extracted from the \texttt{NumObs} and \texttt{UnnObs} databases,
the orbits are from the \texttt{MPCORB} database,
and the pointing from the \texttt{skycov} database.
The asteroids with sufficiently well determined orbits are numerically propagated
to the epoch of observation to determine their celestial coordinates,
nominal apparent magnitude $V$, and apparent velocity $U$.
The celestial coordinates are then matched against the sky coverage data
to determine whether or not an asteroid enters the field of view (FOV) of the survey during that night.
The orbits are propagated numerically by integrating the equations of motion,
using an efficient numerical integrator with variable timestep \citep{1974CeMec..10...35E},
and including gravitational perturbations from the planets,
and then saved at intervals of 200 days.
Then the orbits are interpolated between the two closest 200-days epochs to the observation epoch.
This results in an overall typical error in the position of the asteroids
between 10 and 20 arcsec, 
which is sufficient
for the purposes of this study, and it does not significantly affect the overall outcome.
This choice allows one to reduce by approximately two orders of magnitude the computational costs
related to the orbit propagation, and the level of accuracy allows the inclusion of orbits with 
RUNOFF parameter smaller than 20 arcsec per decade, or MPC orbit uncertainty parameter $\leq 2$
\citep{1978AJ.....83...64M}.

Data are then binned in apparent magnitude $V$ and apparent velocity $U$,
between 14 and 24 with increments of 0.25 in $V$,
and between 0 and 100 arcsec/h with increments of 2 arcsec/h in $U$.
The ratio between detected asteroids $N_\text{obs}$ and total asteroids $N_\text{tot}$ present in the FOV determines
the detection efficiency $\eta$ for each bin:
\[ \eta = \frac{N_\text{obs}}{N_\text{tot}} \addtag \]
where $N_\text{obs}\leq N_\text{tot}$.
Using binomial statistics we can estimate the uncertainty $\sigma_\eta$:
\[ \sigma_\eta = \sqrt{\frac{p(1-p)}{N_\text{tot}}} \addtag \]
which is particularly important here because the bin size is relatively small,
so both $N_\text{obs}$ and $N_\text{tot}$ can have relatively large statistical fluctuations.
The value of the nominal success probability $0<p<1$ of the binomial distribution
is unknown in general for each bin, but we know that it 
tends asymptotically to $N_\text{obs}/N_\text{tot}$ for large $N_\text{tot}$,
and that it must strictly avoid the singular values $p=0$ and $p=1$.
A suitable choice is then
\[ p = \frac{N_\text{obs}+1}{N_\text{tot}+2} \addtag \]
so that the estimate $\eta \pm \sigma_\eta$ is always well defined.

\begin{figure}
\centering
\includegraphics*[angle=270,width=0.90\columnwidth]{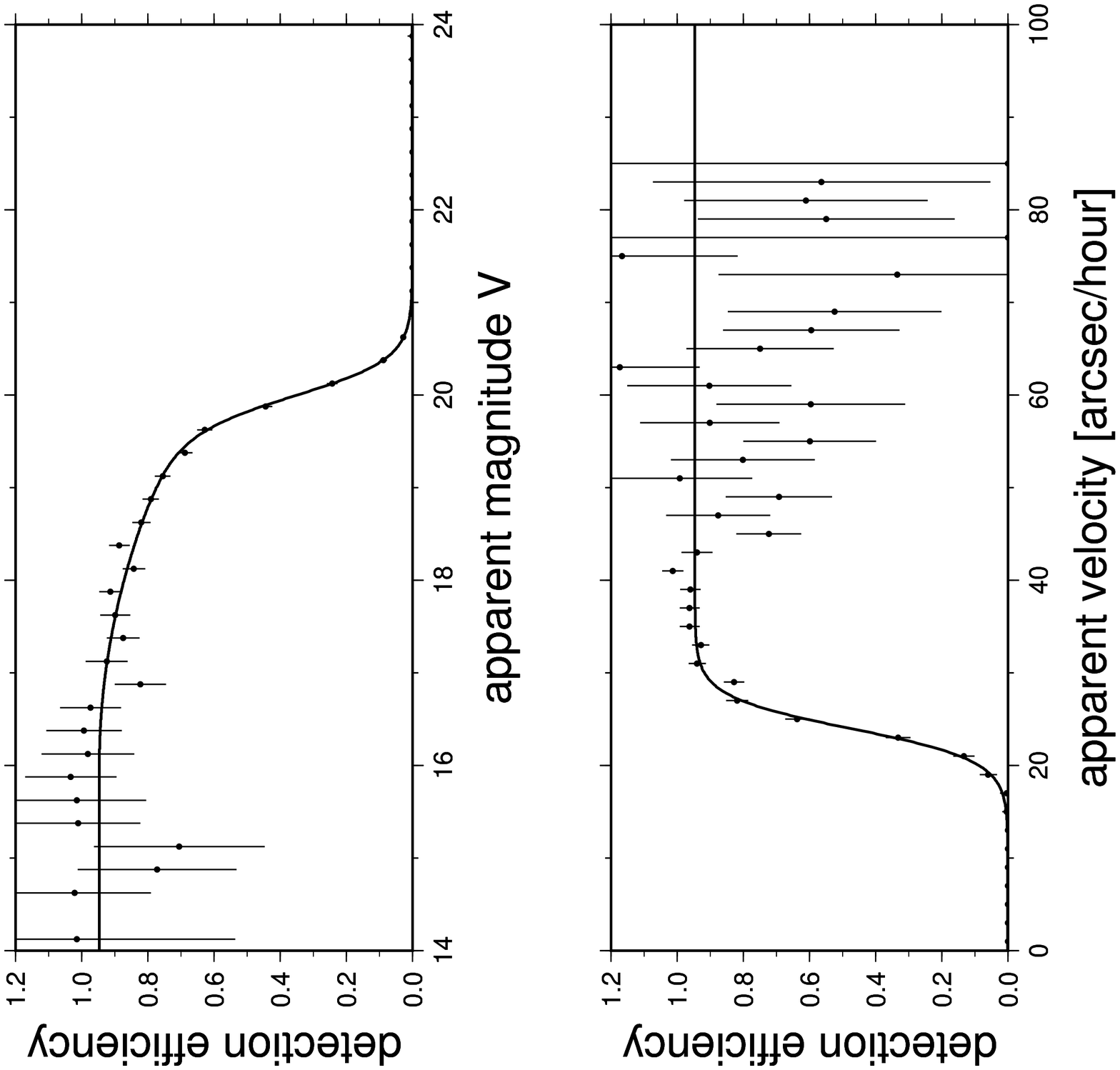}
\caption{
Sample detection efficiency fitting curve as a function of apparent magnitude (top) and apparent velocity (bottom),
along with the data points and related statistical uncertainties.
The fitting parameters are 
$\eta_0=0.95\pm0.02$,
$V_m=20.03\pm0.03$,
$\Delta_V=0.20\pm0.01$,
$q_V=7.1\pm0.5$,
$U_m=24.0\pm0.2$~arcsec/hour,
and 
$\Delta_U=1.7\pm0.1$~arcsec/hour.
The derived limiting magnitude in this example is $V_{50}=19.81\pm0.03$,
while the limiting apparent velocity is $U_{50}=24.2\pm0.2$~arcsec/hour.
}
\label{fig:plot_V_U}
\end{figure}

\begin{figure*}
\centering
\includegraphics*[angle=270,width=0.90\textwidth]{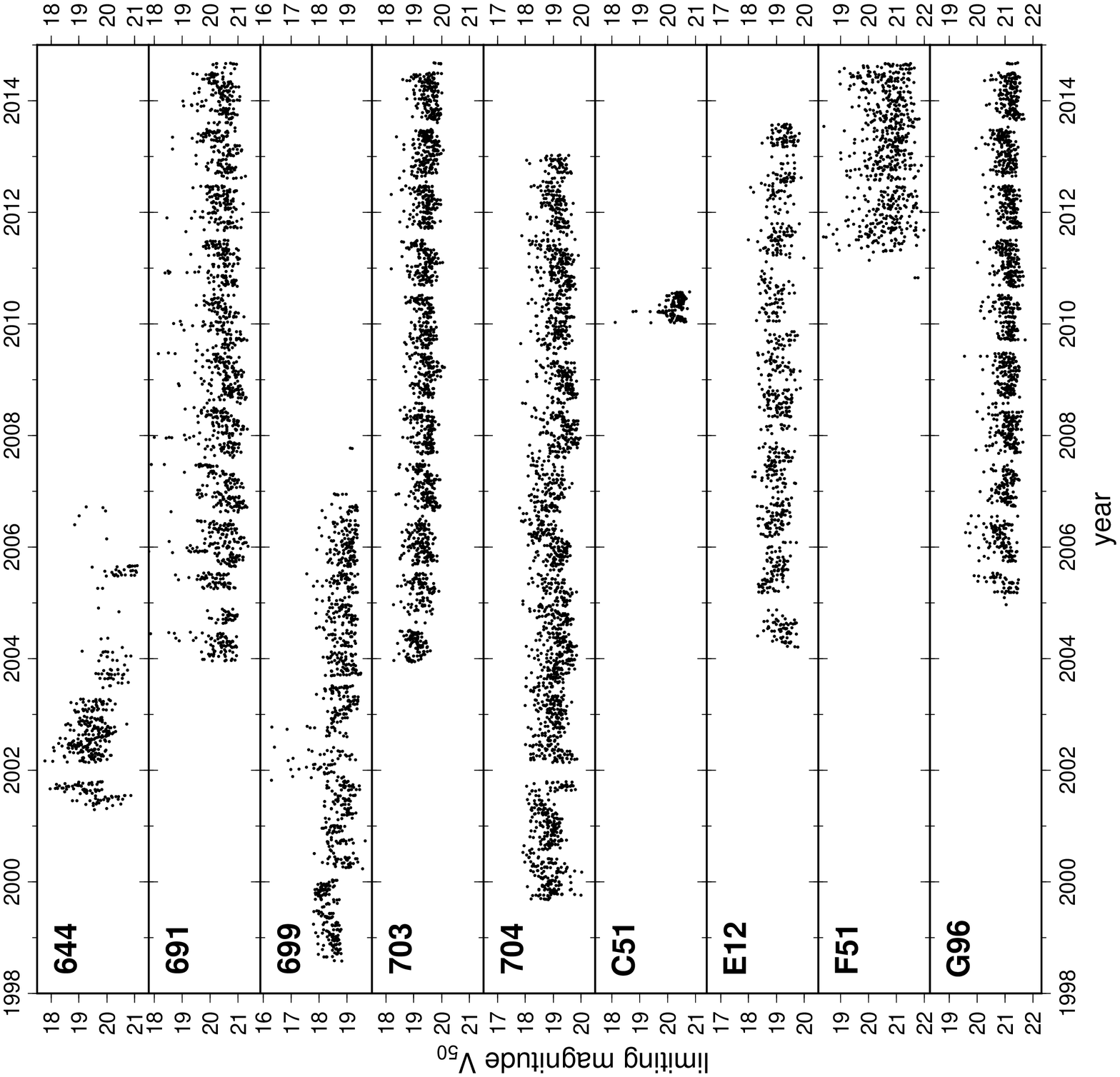}
\caption{
Each point represents the limiting magnitude $V_{50}$ of a single observing night.
The formal statistical uncertainty of data points is not displayed and is typically of the order of few hundredths of a magnitude.
Note how the scale of each plot is the same, only the vertical axis is shifted up or down
in order to include the data points, and this allows to compare directly
the dispersion of the $V_{50}$ for different surveys.
}
\label{fig:grid_V50_scatter}
\end{figure*}

\begin{figure*}
\centering
\includegraphics*[angle=270,width=0.89\textwidth]{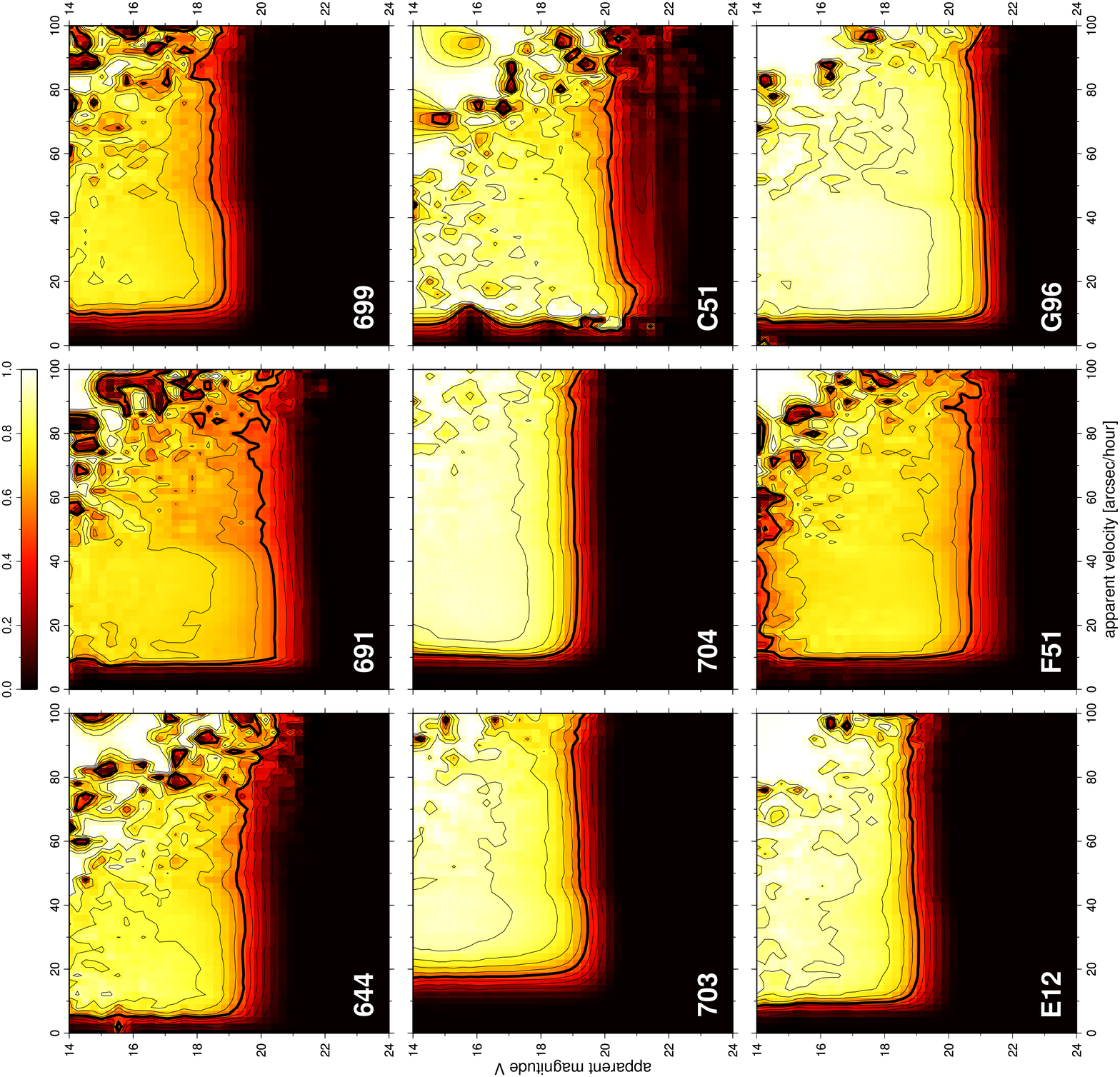}
\caption{Grid plot of the cumulative detection efficiency for each of the nine surveys studied.
Each plot has the same axes range,
and level curves are between 0.1 and 0.9 in increments of 0.1,
with the bold level curve marking the value of 0.5.
}
\label{fig:grid_V_U_data_3x3}
\end{figure*}

As we described in the introduction, there have been
several approaches to fit the detection efficiency in the literature.
Here we choose to analyze its dependence on the apparent magnitude of asteroids
as well as their rate of motion.
The functional dependence on $V$ is adapted from \cite{2009Icar..202..104G},
while the dependence of apparent velocity $U$ is a sigmoid function.
So our choice is to model the detection efficiency as
\[
\eta_\text{fit} = \eta_0 \eta_V \eta_U
\addtag
\label{eq:eta_fit} 
\]
where $0 \leq \eta_0 \leq 1$ is the maximum value of the detection efficiency.
The two other components express the dependence on apparent magnitude $V$:
\[ \eta_V = \frac{\displaystyle 1-\left(\frac{V-V_0}{q_V}\right)^2}{\displaystyle 1+\exp\left(\frac{V-V_m}{\Delta_V}\right)}
\addtag
\label{eq:eta_V} 
\]
and apparent velocity $U$:
\[ \eta_U = \frac{1}{\displaystyle 1+\exp\left(\frac{U_m-U}{\Delta_U}\right)} .
\addtag
\label{eq:eta_U}
\]
The functions are limited between 0 and 1,
and we fix $\eta_V=1$ for $V\le V_0$.
The parameters $V_m$ and $U_m$ are the sigmoid mid-points,
while $\Delta_V$ and $\Delta_U$ are the widths over which the transition occurs.
The confusion effect due to the increasing number of background stars
is taken to be still negligible at $V_0=16$ and is regulated by the $q_V$ parameter.
An example of the fitting of the detection efficiency data
is included in Figure~\ref{fig:plot_V_U}.
Here we define the limiting magnitude $V_{50}$ as the magnitude at which $\eta_\text{fit}=0.50$
and typically $V_{50} \lesssim V_m$ and their difference increases if $\eta_0 < 1$
or if there is a strong quadratic decay (small $q_V$).
For this reason $V_{50}$ is computed numerically, and it is not defined if $\eta_0<0.50$.
Similarly we define $U_{50}$ as the apparent velocity at which $\eta_\text{fit}=0.50$,
and $U_{50} \gtrsim U_m$.
Note that the detection efficiency for $U$ increases with the apparent velocity,
so the limiting apparent velocity $U_{50}$ is a lower limit, not an upper limit.

\section{Results}

The process of determining the detection efficiency curve has been performed for all surveys in Table~\ref{tab:survey_properties}
and for all the nights for which data were available.
Typically a number of asteroids of 100 or more needs to be observed in a single night in order for the fitting process
to converge, and even then there may be several cases when the fit does not converge
due to noisy data, and the night is discarded in that case.
The statistics of nights for which data are available,
and nights for which a good fit was obtained,
is provided in Table~\ref{tab:survey_properties},
and overall we obtained good fits of 10,033 nights.
Data are typically more noisy in apparent velocity, 
especially for apparent velocities above approximately 40 arcsec/hour
where the number of main belt asteroids drops and the statistics with it.
Additionally, there may be nights when the asteroid observing strategy 
tends to sample apparent velocities away from $U_m$ so the data constrain the curve only weakly.
The limiting magnitude for the nights with satisfactory fitting are displayed in Figure~\ref{fig:grid_V50_scatter},
where one can appreciate the period of operation, seasonal and global trends for each survey.
An online archive\footnote{\url{http://orbit.psi.edu/\texttildelow{}tricaric/AsteroidDetectionEfficiency/}}
has been created with individual detailed reports for each night.

An in-depth look at the detection efficiency data accumulated over the full performance period
can confirm the dependence on apparent magnitude and apparent velocity,
as well as reveal subtle structures that would otherwise be lost in the statistical noise of single-night data.
As the plots in Fig.~\ref{fig:grid_V_U_data_3x3} show,
the typical shape of the bold level curve corresponding to a detection efficiency $\eta=0.5$
confirms a posteriori the fact that the fitting function can be separated
into two independent functions depending on $V$ and $U$.
Not all surveys seem to reach detection efficiency close to 1.0, even in the limit of bright asteroids.
The detection efficiency typically shows a uniform front as we move left to right from almost apparently static asteroids to faster ones,
and a modulation when approaching limiting magnitudes, slightly favoring asteroids up to approximately $40$~arcsec/h, which is the range of main belt asteroids.
This effect can be somehow observed in data from individual nights, see e.g.~Figure~\ref{fig:plot_V_U},
but data from single nights are typically insufficient to obtain statistically robust fits of this effect.
Also, the noise in the data increases markedly above $40$~arcsec/h, that is due principally to the statistics.

\begin{figure}
\centering
\includegraphics*[angle=270,width=0.7\columnwidth]{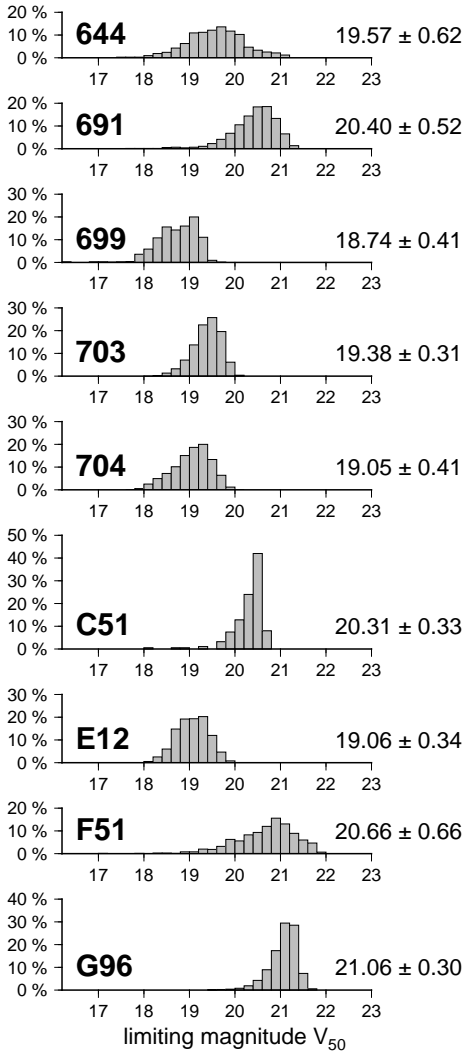}
\caption{Histograms of the observing nights at a given limiting magnitude $V_{50}$.
The average and standard deviation is marked next to each distribution.
}
\label{fig:grid_V50_histo}
\end{figure}

The distribution of nightly limiting magnitudes is provided in Figure~\ref{fig:grid_V50_histo},
along with the basic statistics.
It is interesting to note that for most surveys the distribution clearly shows a slow rise and then a quick drop
as we move left to right towards increasing $V_{50}$ values, resulting in skewed distributions.
Our average $V_{50}$ are in very good agreement with the average $V_\text{lim}$
for 703 (19.44) and G96 (21.15) in \cite{Jedicke_CSS},
within $0.1$ magnitudes, where their $V_\text{lim}$ is defined as the magnitude at which the efficiency drops to $\eta_0/2$,
leading to slightly higher values than our $V_{50}$ defined as the magnitude at which $\eta_\text{fit}=0.5$.

\begin{figure}
\centering
\includegraphics*[angle=270,width=0.659\columnwidth]{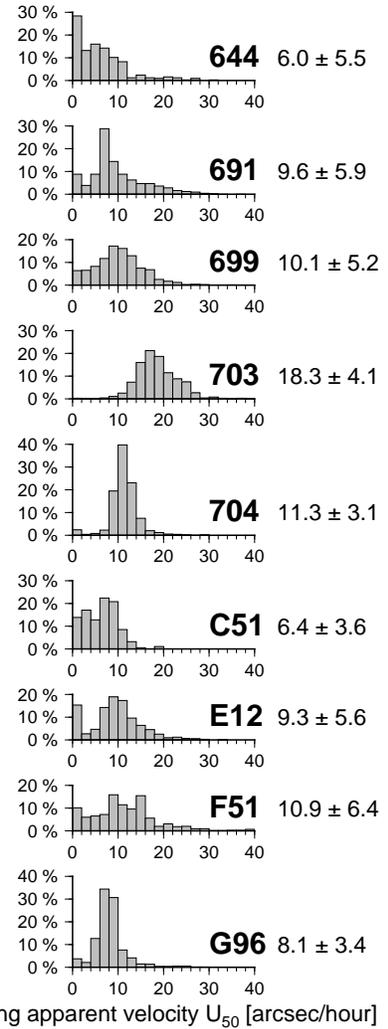}
\caption{Histograms of the observing nights at a given limiting apparent velocity $U_{50}$.
The average and standard deviation is marked next to each distribution.
}
\label{fig:grid_U50_histo}
\end{figure}

The distribution of nightly limiting apparent velocities is provided in Figure~\ref{fig:grid_U50_histo}.
Here we see two groups of distributions: those compact at values higher than zero, and those with a significant component near zero.
Low $U_{50}$ values can indicate search strategies with an observing cadence re-visiting the same FOV over a long time period,
or can be a sign of reduced data resolution at low $U$ values and thus provide only weak constraints on $U_{50}$.

\begin{figure}
\centering
\includegraphics*[angle=270,width=0.624\columnwidth]{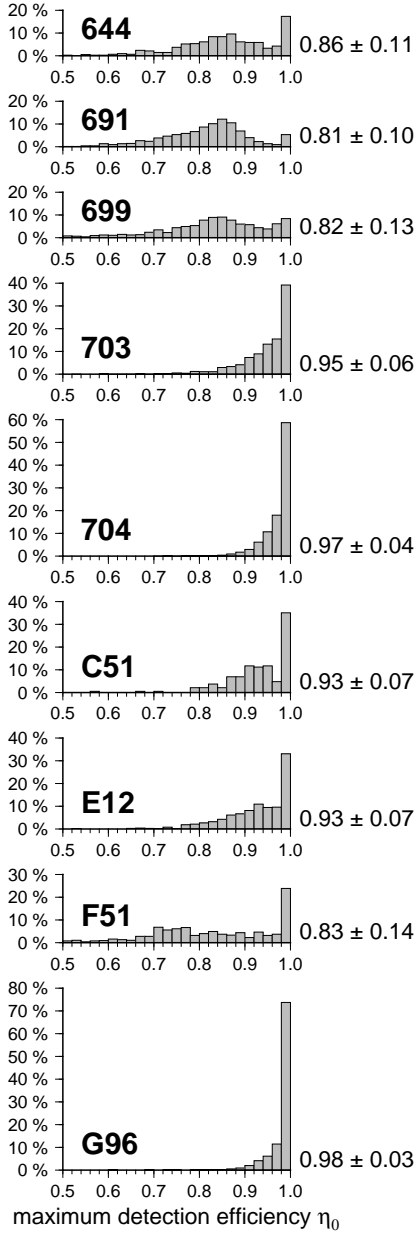}
\caption{Histograms of the observing nights at a given maximum detection efficiency $\eta_0$.
The average and standard deviation is marked next to each distribution.
}
\label{fig:grid_eta0_histo}
\end{figure}

In Figure~\ref{fig:grid_eta0_histo} we have the distribution of maximum detection efficiencies $\eta_0$.
Here too we can see two groups of distributions: those strongly peaked towards $\eta_0=1$,
and those with a significant fraction of nights below 1.0.
The value of $\eta_0$ for NEOWISE C51 is in generally good agreement with \cite{2011ApJ...743..156M}.
However our values are significantly higher than the values for 703 (0.73) and G96 (0.88) provided in \cite{Jedicke_CSS}.
The lower $\eta_0$ values in \cite{Jedicke_CSS} are probably due to them not including
the apparent velocity in the fit of the detection efficiency.
Since both 703 and G96 have 
$U_{50}$ distribution values well above zero (Figure~\ref{fig:grid_U50_histo}), this effect could be quite substantial in decreasing the apparent value of $\eta_0$.
On the other hand, this effect is minimal for C51 as their $U_{50}$ distribution is very close to zero.
The lack of a slow quadratic decay term in the apparent magnitude fitting function can also contribute to underestimating the $\eta_0$.

\begin{figure}
\centering
\includegraphics*[width=\columnwidth]{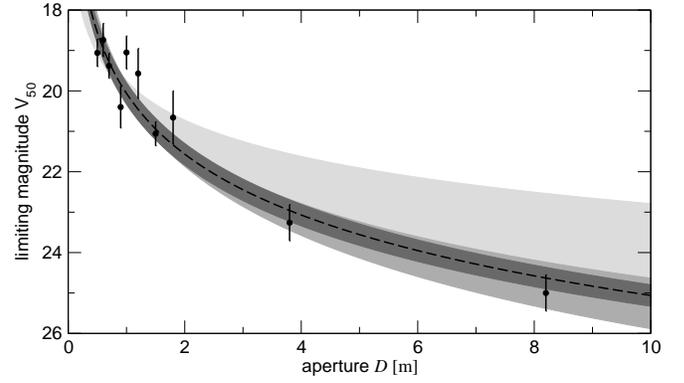}
\caption{
Limiting magnitude $V_{50}$ of ground-based asteroid surveys as a function of aperture $D$.
The data points in the left group ($D<2$~m) are from this work,
while the point at $D=3.8$~m is based on \cite{2009Icar..202..104G} with data from the KPNO Mayall telescope,
and the point at $D=8.2$~m is based on \cite{2007P&SS...55.1113Y} with data from the NAOJ Subaru telescope.
The fitting dashed line is $V_{50} = 20.06 + 5 \log_{10} D$ with $D$ in meters,
and the gray regions mark the $2\sigma$ uncertainty of the fit under different assumptions (see main text).
}
\label{fig:V50_chisq}
\end{figure}

The limiting magnitude $V_{50}$ at which an asteroid survey operates depends primarily on the aperture $D$ of the telescope, 
with corrections from additional factors such as exposure time, weather conditions, and search strategies.
Here we have an opportunity to look at the general trend of $V_{50}$ versus $D$
having characterized each survey using exactly the same approach based on the same data sources.
For consistency we limit this analysis to the 8 ground-based surveys included in this study,
and then discuss the comparison with the only one space-based, NEOWISE.
In Figure~\ref{fig:V50_chisq} we show the values of the limiting magnitude $V_{50}$ versus aperture $D$.
When restricted to these 8 ground-based surveys, the best fitting function is 
\[ V_{50} = (19.95 \pm 0.11) + (4.12 \pm 0.60) \log_{10}D \addtag \]
where $D$ is in meters
and the uncertainties quoted are $1\sigma$.
In Figure~\ref{fig:V50_chisq} the $2\sigma$ range of solutions,
representing the 95\% confidence region, is represented by the light gray area.
The reduced chi-squared is $\tilde{\chi}^2 = 1.75$,
relatively high and indicates the difficulty to constrain the factor of the logarithm,
mostly because of the limited range in aperture $0.5 \le D \le 1.8$~meters in the input data.
To extend this aperture range we can include results from two large-aperture telescopes,
the Kitt Peak National Observatory (KPNO) Mayall telescope ($D=3.8$~m),
and the National Astronomical Observatory of Japan (NAOJ) Subaru telescope ($D=8.2$~m).
\cite{2009Icar..202..104G} searched for asteroid for 6 nights in 2001 at the Mayall telescope,
and determined the detection efficiency for each night by injecting synthetic asteroids in their images
and then reprocessing the images using the same pipeline used for real asteroids.
When adapted to the conventions used in this manuscript,
and after applying the standard filter correction $V-R=0.4$,
their overall limiting magnitude is estimated as $V_{50} = 23.26 \pm 0.45$.
Similarly \cite{2007P&SS...55.1113Y} performed asteroids search at the Subaru telescope on a single night also in 2001,
and used synthetic asteroids to estimate the detection efficiency,
and their overall limiting magnitude is estimated as $V_{50} = 25.00 \pm 0.45$.
Note that in both cases the nominal $1\sigma$ uncertainty of $0.45$ magnitudes
is estimated using the average uncertainty of the other 8 ground-based surveys.
The best fit for all 10 ground-based surveys is
\[ V_{50} = (20.04 \pm 0.12) + (5.23 \pm 0.33) \log_{10}D \addtag \]
with a reduced chi-squared of
$\tilde{\chi}^2 = 1.60$,
and represented by the intermediate gray area in Figure~\ref{fig:V50_chisq}.
It is clear how these additional points help constrain the limiting magnitude at large aperture,
with the logarithm factor now within less than $1\sigma$ from the theoretical value of 5.
If we fix the logarithm factor to be exactly 5,
we obtain
\[ V_{50} = (20.06 \pm 0.15) + 5 \log_{10}D \addtag \label{eq:fit10log5} \]
with a reduced chi-squared of
$\tilde{\chi}^2 = 1.45$
and represented by the dark gray area in Figure~\ref{fig:V50_chisq}.
This last fit of Eq.~\eqref{eq:fit10log5} should be considered the most robust of the three
to estimate the expected performance of upcoming ground-based surveys.
As Figure~\ref{fig:V50_chisq} shows, small deviations are possible 
and due most likely to the details of the asteroid search operations.
Note that when large aperture telescopes search Kuiper Belt Objects (KBOs)
the limiting magnitude can improve over Eq.~\eqref{eq:fit10log5} by one magnitude or more,
as the low apparent velocities allow longer exposure times
\citep{2001AJ....122..457T,2015arXiv151102895B}.
When considering infrared space surveys, 
the single NEOWISE data point of 
$V_{50} = 20.31 \pm 0.33$ at an aperture $D=0.4$~m
allows us to obtain trivially the fitting curve
\[ V_{50} = (22.30 \pm 0.33) + 5 \log_{10}D \addtag \]
where the statistical uncertainty is simply that of the input data,
and the logarithm factor is fixed at its theoretical value.
This indicates an improvement over ground-based optical surveys of approximately two magnitudes.

\section{Conclusions}

We have presented a method to characterize the detection efficiency
of asteroid surveys, and applied it to nine of the largest surveys active over the past two decades.
The input data maintained by the Minor Planet Center are freely available for all the
surveys considered, and have been uniformly processed by our software pipeline,
and this facilitates an unbiased characterization and aids in the goal
of having the largest statistical significance.
All surveys show large dispersions in limiting magnitude, limiting apparent velocity, and maximum efficiency over
short periods, which underlines the importance of accurately following and modeling these parameters.
This set of ten thousand well characterized observing nights
represents a solid base towards producing debiased asteroid population studies.

The dependence of the limiting magnitude on the aperture
can deviate significantly from the nominal fitting curve,
depending on the asteroid search strategy and observing conditions,
but overall our estimate of Eq.~\eqref{eq:fit10log5} should represent a solid 
first-order expected limiting magnitude for current and future asteroid surveys.

Modeling the detection efficiency as a function of apparent magnitude and apparent velocity
on a nightly basis generated solid fits. Using only the apparent magnitude
can lead to underestimating the overall efficiency for surveys which are significantly insensitive to apparently slow asteroids.
More complex fitting functions could be tested over much longer periods than nightly in order to accumulate larger statistics,
and could model the slow decay for fast moving objects, as well as dependence on the airmass or galactic latitude.

\acknowledgments

This research was supported by the NASA Near Earth Object Observations (NEOO) Program,
grant NNX13AQ43G.
Surveys are gratefully acknowledged for making available the data used in this study.
Gareth Williams and the Minor Planet Center are thanked for curating and maintaining the data repository.
Steve Chesley and the other anonymous referee are gratefully acknowledged for prompt and constructive reviews.

\end{document}